\documentclass{article}
\usepackage{verbatim, url, graphicx, times}
\usepackage[bf]{subfigure}
\usepackage[bf]{caption}
\usepackage[utf8]{inputenc}

\begin{document}
\title{%
Scilab and SIP for Image Processing
}
\author{Ricardo Fabbri$^{1,2,3}$\and Odemir Martinez Bruno$^{1,2}$ \and Luciano da
Fontoura Costa$^{2}$\\[10pt]
$^1$Institute of Mathematical and Computer Sciences\\
University of S\~{a}o Paulo, Brazil\\
\\
$^2$Physics Institute of S\~{a}o Carlos\\
University of S\~{a}o Paulo, Brazil\\
\\
$^3$Polytechnic Institute\\
State University of Rio de Janeiro, Brazil\\
\\
\small{Contacts:}\\ 
\small{rfabbri@iprj.uerj.br}\\ 
\small{bruno@ifsc.usp.br}\\
\small{lfcosta@ifsc.usp.br}
}

\date{\today}
\maketitle
\begin{abstract}
This paper is an overview of Image Processing and Analysis using
Scilab, a free prototyping environment for numerical calculations
similar to Matlab. We demonstrate the capabilities of SIP -- the
Scilab Image Processing Toolbox -- which extends Scilab with many
functions to read and write images in over 100 major file formats,
including PNG, JPEG, BMP, and TIFF. It also provides routines for
image filtering, edge detection, blurring, segmentation, shape
analysis, and image recognition.  Basic directions to install
Scilab and SIP are given, and also a mini-tutorial on Scilab.
Three practical examples of image analysis are presented, in
increasing degrees of complexity,  showing how advanced image
analysis techniques seem uncomplicated in this environment.
\end{abstract}

\section{Introduction}~\label{Sec:intro}
Image Processing and Analysis are fields of Computer Science whose
objective is to enhance digital images and extract information
from them~\cite{gonzalez,parker:1997,Costa:2001}. This allows automatic or
semi-automatic identification, classification or characterization of
objects and patterns. Some interesting applications are biometry
systems (e.g. fingerprint and iris recognition), satellite and microscope
image analysis, diagnosis from medical images, and more. Another
example is special effect filtering, usually found in image
manipulation software such as the \textsc{gimp}~\cite{Gimp} or Photoshop.

A remarkable characteristic of this field is its overwhelming
complexity. Many image processing techniques are based on
sophisticated mathematical and computational concepts, such as
the Fourier Transform. All this is combined with deep biological,
psychological and probabilistic principles underlying the way
animals identify and recognize objects.  This makes implementation a
challenging (frequently tedious) task , specially with traditional
languages such as C or Fortran.  As a consequence, the area had
been somewhat limited to a small community of scientists and
programming experts.

The difficulties involving the development of software in general
and the advances in hardware power gave rise to a quite recent trend
in programing~\cite{Osterhout:1998} -- prototyping with interpreted,
easy-to-use languages, such as Python, Perl, Lua, and Tcl/Tk. Different
algorithms and solutions for a given problem are easily developed
and validated using these environments. When a solution finally works
as expected, the final application may be written in a traditional
compiled language, such as C.  Many programmers also embed an
interpreter in their final application, to reuse the large amount
of functionality available in these environments without having to
rewrite it all.

Following this trend, the inherent complexity of scientific
applications called for specific scripting technologies. Among
them is the popular (and rather expensive) Matlab~\cite{matlab},
very widely used in science, engineering,
and the industry.  Fortunately there are free software alternatives such as
Octave~\cite{octave}, numerical Python~\cite{numpy}, and
Scilab~\cite{Scilaburl} --
the one we will be talking about in this paper.  Features common
to all these numerical prototyping environments are convenient
matrix manipulation enforced by a suitable language, as well as
tools for scientific visualization, debugging, and a great ammount of
easy-to-use libraries. They may be used interactively or programmed
from a separate file (the `script' or 'macro'). In particular, they enable
development of Image Processing applications with much less burden
to the programmer. Hence, these tools have in fact been the standard
prototyping solutions in the field.

Scilab is a free software created at INRIA--France for prototyping
and numerical processing. It is much like Matlab, and already has a
rich set of functionalities. Currently in version 5.4, Scilab have
been widely used in Unix-like systems. It has been adopted in many
Universities and companies around the world. Given its features to
assist numerical programming, as we just said, it has tremendous
potential to be used for image processing, both educationally and as
a prototyping language to develop and test solutions.  
As far as we know, currently neither Octave or Python match the richness and
plain simplicity of scilab functionality for scientific computing and
engineering.

The following
listing summarizes Scilab's features and how they compare to Matlab's:

\begin{itemize}
\item \textbf{Expressive programming language} with similar
syntax to Matlab's, although not identical. It presents many
sophisticated construct improvements over Matlab, having relatively
small drawbacks. Moreover, operations are expressed in much the same
way they are usually written mathematically, which enforces simplicity
and organization.

\item \textbf{Elaborate data structures:} polynomial, complex and
string matrices, lists, linear systems, records (tlists), and graphs.
In these subjects there are significant improvements over Matlab.

\item \textbf{Graphical and Scientific Visualization library:} 2D, 3D, and
animations. Even though the graphical package is not as stable as Matlab,
it is useful and complete. It has been continually improved and optimized.
The GUI itself has now been written in Java, and looks and feels very similar --
if not better -- than Matlab.

\item \textbf{Modular structure:} Scilab allows easy interfacing to
Fortran and C using dynamic linking.

\item \textbf{Many sophisticated libraries:}
\begin{itemize}
   \item \textbf{Linear Algebra}, including sparse matrices.
   \item \textbf{Signal Processing} with a comprehensive manual;
   \item \textbf{Statistics}
   \item \textbf{Scicos}, a dynamic systems modeler and simulator,  similar to
   Matlab's Simulink.
   \item \textbf{Metanet}: graphs and networks, including a graphical editor
   \item \textbf{Control}
   \item \textbf{Simulation and Modeling}: ODE solver, etc.
   \item \textbf{Optimization} (differential and non-differential)
   \item \textbf{Interface with symbolic algebra} packages such as Maple
   \item \textbf{Interface with Tcl/Tk}, useful for building GUI's
   \item \textbf{LMI optimization} (Linear Matrix Inequalities)
   \item \textbf{Many contributed libraries:} genetic
   algorithms, neural networks, plotting enhancements, other numerical
   methods, and, of course, image processing.
\end{itemize}
\end{itemize}

Scilab's potential for imaging and its maturity under GNU/Linux systems motivated
the creation of SIP -- the Scilab Image Processing
Toolbox~\cite{Fabbri}. SIP
extends Scilab with many routines to deal with image files and analyze
them. Examples of its main features are:
\begin{itemize}
\item \textbf{Functions to read and write many image formats.} Almost 100
major formats are supported, including JPEG, PNG, BMP, GIF, and TIFF
-- due to the underlying ImageMagick library~\cite{ImageMagickurl}.

\item \textbf{Filters:} Gaussian blurring, median, Laplacian,
artistic effects, de-noising by min-max curvature flow.

\item \textbf{Edge detection:} Sobel, Fourier derivatives, Canny.

\item \textbf{Geometric transforms:} rotation, zoom, shearing, and
general affine.

\item \textbf{Image segmentation:} watersheds, adaptive thresholding.

\item \textbf{Mathematical morphology:} dilation, erosion, thinning, etc., using
state-of-the-art Euclidean algorithms for circular structuring elements.

\item \textbf{Shape analysis:} perimeter, border tracking, state-of-the-art fast
skeletonization with multiscale pruning, curvature, state-of-the-art Euclidean
distance transforms~\cite{Fabbri:EDT:survey:ACMCSurv08} (including fast distance transforms up to a given
distance), fractal dimension.

\item \textbf{Enhancement:} histogram equalization, contrast manipulation.

\item \textbf{Other operators:} Hough transform, noise generation,
image display, interferometry operations (e.g. phase unwrapping), etc.

\item \textbf{It's fast:} heavier operations are implemented
in C language, allowing for efficiency. Many operations use
state-of-the-art efficient algorithms, some of which are so new
or advanced as to be absent in most if not all proprietary image
analysis software.

\item \textbf{It's well documented:} Every function has a help page
containing its reference and a practical example.

\item \textbf{It's free:} General Public License (GPL).

\item \textbf{It's easy:} Useful for fast programming and experimentation.
\end{itemize}

All these tools can be easily used interactively or from scilab
scripts, even though some would extremely complex to implement
from scratch.  Thus, a programmer without specialized knowledge
may readily experiment with different techniques to solve his
particular problem.

In this paper we show how to use SIP/Scilab and demonstrate their
benefits through a hands-on approach. The examples will show
how complex operations seem approachable under the expressiveness of the
environment. Rather than worrying about a single imaging operator,
the focus is shifted to the concatenation of various operators into
a useful solution.

We start with brief directions on how to install Scilab under Linux
and present a tutorial introduction to it. We then show how to
install SIP and use it in three practical problems from Image
Analysis and Processing: edge detection, OCR, and separation of
overlapping blood cells. The last two illustrate techniques like
noise removal, skeletonization, and object segmentation.

\section{Customized Scilab Installation}
To build a special version of Scilab by compiling the
sourcecode, the user must have all development packages installed
in the system (GCC and gfortran compilers, and so on).
After downloading and unpacking the source tarball, one types
the following inside Scilab's directory:

\begin{verbatim}
      $  ./configure
      $  make all
      $  make install
\end{verbatim}

To start scilab: 
\begin{verbatim}
      $ scilab
\end{verbatim}

One may enter commands in the scilab window, as shown
in the following section. There is an important add-on toolbox
called ``Enrico'' which greatly enhances Scilab's graphical
capabilities. It is very simple to install, and may be downloaded at: 
http://www.weizmann.ac.il/~fesegre/scistuff.html

The user must unpack it, start Scilab, then type:
\begin{verbatim}
--> exec '/path/to/ENRICO/builder.sce';
\end{verbatim}

This compiles the toolbox. The user may now load it:
\begin{verbatim}
--> exec '/path/to/ENRICO/loader.sce';
\end{verbatim}
This command may be placed in the ``.scilab'' file to load it
automatically whenever one enters Scilab.  In section~\ref{Sec:blobs}
we show how to render a 3D surface using this toolbox (see
figure~\ref{Fig:cells:dt}).

\section{A Mini Tutorial on Scilab}

In this section we present some examples to illustrate the ease of
doing programming and calculations with Scilab. We hope to attest its
particularly great potential for programming and for developing image
analysis solutions. Initially, we use Scilab in
interactive mode. Commands are typed and executed directly in
Scilab's shell. 

A vector is created like this:
\begin{verbatim}
--> a = [1 2 3 4]
 a  =

!   1.    2.    3.    4. !
\end{verbatim}

User commands are typed at the `\texttt{-->}' prompt; Scilab's
responses have no preceding prompts.
To define a column vector, we use commas:
\begin{verbatim}
--> a = [1; 2; 3; 4]
 a  =

!   1. !
!   2. !
!   3. !
!   4. !
\end{verbatim}

This same vector could have been created using the `:' (colon) operator:
\begin{verbatim}
--> a = 1:4
 a  =

!   1.    2.    3.    4. !
\end{verbatim}

Matrices are also very simple to define, as shown below:
\begin{verbatim}
--> a = [1 2; 3 4]
 a  =

!   1.    2. !
!   3.    4. !
\end{verbatim}

Scilab have convenient operators for naturally manipulating matrices.
For instance, to obtain the trasposed matrix of $a$:
\begin{verbatim}
--> b = a'
 b  =

!   1.    3. !
!   2.    4. !
\end{verbatim}

In this other example, we perform matrix multiplication of $a$ and $b$:
\begin{verbatim}
--> c = a*b
 c  =

!   5.     11. !
!   11.    25. !
\end{verbatim}

Scilab also has many functions readily available for usage. Below we
calculate the determinant of $c$:
\begin{verbatim}
--> d = det(c)
 d  =

    4.
\end{verbatim}

In addition to the simplicity of operating with arrays, there are many
other mathematical functions, as well as visualization facilities.
We will now show this by a simple example about the Fourier transform.

\begin{figure}[htb]
\centering
\scalebox{0.3}{\includegraphics{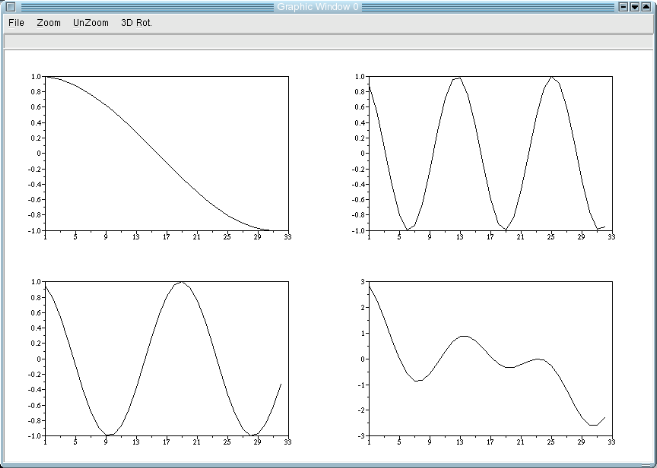}}
\caption{Plots of vectors \texttt{c1}, \texttt{c2}, \texttt{c3}, and
\texttt{c}}\label{Fig:PlotSins}
\end{figure}

As the first step, we create a vector by summing three sampled
cosines:
\begin{verbatim}
--> a  = 1:32;  // an array with 32 elements (1 to 32).
--> c1 = cos(a/10); 
--> c2 = cos(a/2);
--> c3 = cos(a/6);
--> c  = c1 + c2 + c3;
\end{verbatim}

Now we plot the vectors. We'll use a simple function called
\texttt{plot} for this, together with the command \texttt{subplot}
to divide the graphical window. The commands below yield
figure~\ref{Fig:PlotSins}.
\begin{verbatim}
--> subplot(2,2,1), plot(c1);
--> subplot(2,2,2), plot(c2);
--> subplot(2,2,3), plot(c3);
--> subplot(2,2,4), plot(c);
\end{verbatim}

\begin{figure}[htb]
\centering
\scalebox{0.35}{\includegraphics{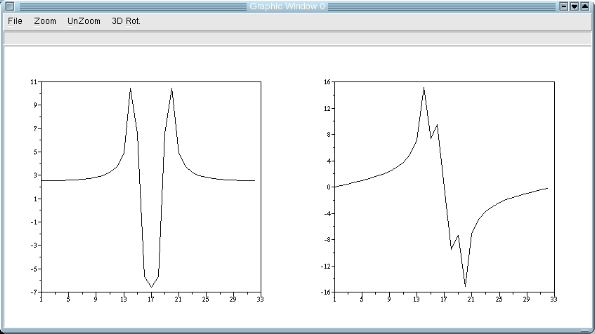}}
\caption{Real (left) and imaginary part (right) of the complex vector
\texttt{hc}}\label{Fig:RealImag}
\end{figure}

We will now apply the Fourier transform to the vector \texttt{c},
using function \texttt{fft} from the signal processing toolbox.
Notice how simple it is:
\begin{verbatim}
--> hc = fft(c,-1);
\end{verbatim}

Observe that the second parameter indicates the sign of the
transform's exponent -- it's not the inverse transform. Vector
\texttt{hc} is made of complex numbers representing the result of
the \texttt{fft} applied to \texttt{c}.  To see the results of the
transform, lets use \texttt{plot}. Since a complex vector is composed
of real and imaginary parts, we use the functions \texttt{real}
and \texttt{imag} to return the real and imaginary
components of the array. Figure~\ref{Fig:RealImag} shows the plots of
the real and imaginary parts of \texttt{hc}.
\begin{verbatim}
--> subplot(1,2,1), plot(real(fftshift(hc)));
--> subplot(1,2,2), plot(imag(fftshift(hc)));
\end{verbatim}

To finish this example, we perform the inverse FFT and compare the
result with the original vector:
\begin{verbatim}
--> fc = fft(hc,1);
--> subplot(1,2,1), plot(fc);
--> subplot(1,2,2), plot(c);
\end{verbatim}

\begin{figure}[htb]
\centering
\scalebox{0.318}{\includegraphics{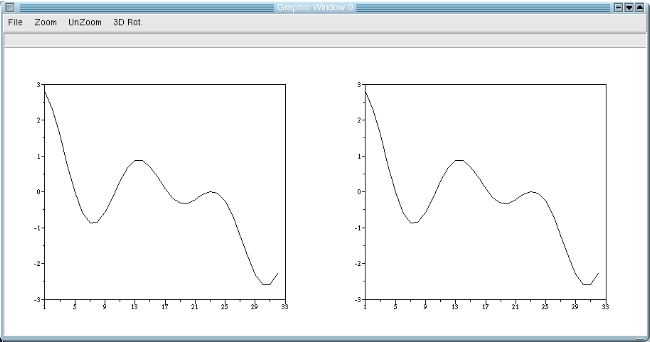}}
\caption{%
Comparation between the inverse FFT and the original vector.
}\label{Fig:ifft}
\end{figure}

\section{Installing the SIP Toolbox}
The previous example shows how Scilab is really an easy and powerful
for numerical programming and array manipulation.  It would be great
to use Scilab's power and convenience to process digital images,
since any image is an array of pixels.  To do this, the SIP (Scilab Image
Processing toolbox) was created (cf. section~\ref{Sec:intro}). It
extends Scilab with functions to read and write images stored in
almost any popular format (over 80), including PNG, JPEG, BMP, and
TIFF. SIP also provides a rich set of ready-to-use operations for
digital image enhancement and analysis, useful for many practical
problems.

It is straightforward to install SIP under Linux.  Binary packages
are provided for Intel pentium and compatible processors. To install
them, the user first downloads and unpacks the tarball. Inside Scilab,
one types:
\begin{verbatim}
$  exec /path/to/sip/loader.sce;
\end{verbatim}
This loads the toolbox. This command may be placed in the file
\texttt{.scilab} under the user's home directory to automatically load
the toolbox whenever Scilab is entered.

For non-Intel architectures, SIP must be installed from source,
which is also necessary to install the latest development
version~\cite{Fabbridev}.
This requires first installing the dependencies, in this order:
ImageMagick, Animal~\cite{Animal}, OpenCV (optional), and
finally SIP proper (cf. the links section). Once downloaded and
unpacked, they are installed like this:

ImageMagick (responsible for the file format support):
\begin{verbatim}
      $  ./configure --enable-shared
      $  make
      $  make install
      $  ldconfig
\end{verbatim}
The latter command updates the system cache of libraries.

Animal (core C routines):
\begin{verbatim}
      $  ./configure
      $  make
      $  make install
      $  ldconfig
\end{verbatim}

SIP (the toolbox proper):
\begin{verbatim}
      $  ./configure
      $  make
      $  make install
\end{verbatim}

SIP is by default installed under Scilab's directory at
`\texttt{contrib/sip}'. Now the toolbox may be loaded just like
the binary version. Alternatively, one may type: 
\begin{verbatim}
      $  make autoload
\end{verbatim}

Which will write to the file \texttt{.scilab} and make the toolbox load
automatically every time you enter Scilab. More details are
found in the file \texttt{INSTALL.txt} included in the SIP sourcecode
package.~\footnote{If anything goes wrong, the user is encouraged
to subscribe to the SIP toolbox user's email list.}

To check if the installation went well, the user steps through SIP's demos:
\begin{verbatim}
$  exec(SIPDEMO);
\end{verbatim}
Which shows the basics of image structure and manipulation.

\section{Fundametal Image Operations}

To read an image into a Scilab variable, one must type, for example:
\begin{verbatim}
--> im = imread('myimage.png');
\end{verbatim}

for grayscale or truecolor images.  Or

\begin{verbatim}
--> [im,map] = imread('myimage.png');
\end{verbatim}

for pseudocolor images. The file format is automatically
detected. From now on the image is in memory, stored as an array of
pixels. We call \emph{grayscale} those images represented as a 2D array
with values from $0$ to $2^{16} - 1$ $(65535)$. \emph{Truecolor}
images are made of three grayscale images, one for each channel
(Red, Green and Blue), represented in a 3D array $(N\times
M\times3)$. \emph{Pseudocolor} images, also known as paletted or indexed
images, are made of two arrays: index and map. The map is an
$(N\times3)$ array with $N$ colors, while the index array is a
2D array whose entries point to a color in the colormap. Further
details are in the introductory demo that comes with SIP.

To show a truecolor or grayscale image:
\begin{verbatim}
--> imshow(im);
\end{verbatim}

For paletted images:
\begin{verbatim}
--> imshow(im,map);
\end{verbatim}

To write an image:
\begin{verbatim}
--> imwrite(im,'foo.jpeg');
\end{verbatim}

or
\begin{verbatim}
--> imwrite(im, map,'foo.jpeg');
\end{verbatim}

\begin{figure}[htb]
\centering
   \subfigure[]{ %
      \label{Fig:arara}%
      \begin{minipage}[c]{0.5\linewidth}%
         \centering
         \scalebox{0.5}{\includegraphics{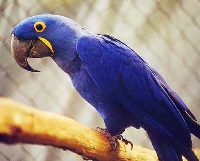}}\\[2mm]
      \end{minipage}}%
   \subfigure[]{ %
      \label{Fig:edges}%
      \begin{minipage}[c]{0.5\linewidth}%
         \centering
         \scalebox{0.5}{\includegraphics{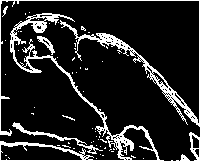}}
      \end{minipage}}\\
   \subfigure[]{%
      \label{Fig:rotated}%
      \begin{minipage}[c]{0.3\linewidth}%
      \centering
         \scalebox{0.5}{\includegraphics{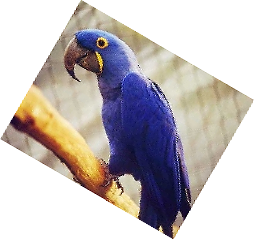}}\\[2mm]
      \end{minipage}}
\caption{%
Introductory example: (a) Read an image of an amazon bird, (b)
extract edges from the image, and (c) rotate the image.
}\label{Fig:operations}
\end{figure}

The next example is about edge detection, one of the most common
operations in image processing~\cite{gonzalez}. First, let us read an image:

\begin{verbatim}
--> a = imread('arara.jpg');
--> imshow(a);
--> a = im2gray(a);       // converts to grayscale
\end{verbatim}
The image is shown in figure~\ref{Fig:arara}.

To find the edges, we use the "edge" command:
\begin{verbatim}
--> b = edge(a,'sobel',0.15);   
--> imshow(b,2);
\end{verbatim}
The parameters means to find edges using the Sobel filter and a 
threshold at $15\%$ intensity level. Figure~\ref{Fig:edges} shows the
result of this technique.

Another common operation with images is rotation and scaling.
The \texttt{mogrify} command wraps many functionalities from the
ImageMagick library, and may be used to rotate the image.  We must
first put the image in the $0$ -- $2^{16} - 1$ range, because mogrify
and other routines such as imwrite assumes 16bit graylevels. We
just multiply the image by $257$, since $255 * 257 = 2^{16}$:
\begin{verbatim}
--> a = a*257;   // normalizes image to 0-65535 range
--> c = mogrify(a, '-rotate 30');   // rotates 30 degrees
--> imshow(c, []);  // "[]" stands for  rescaled 
\end{verbatim}

Figure~\ref{Fig:rotated} shows the color image rotated in 30 degrees
clockwise. To rotate the color image, the same steps may be repeated
to each image channel.

\section{Example: OCR}
Let us now move on to more interesting operations. Figure~\ref{Fig:ocr:a}
shows an image of the letter A. To follow this example, 
the input image can be obtained from the following address:
\begin{center}
  \texttt{http://siptoolbox.sourceforge.net/tutorial/images/a.png}
\end{center}

Read the image with the following commands:
\begin{verbatim}
 a = imread('a.png');
 a = im2gray(a);    // converts to grayscale
 imshow(a);
\end{verbatim}
\begin{figure}[htb]
\centering
   \subfigure[]{ %
      \label{Fig:ocr:a}%
      \begin{minipage}[c]{0.3\linewidth}%
         \centering
         \scalebox{1.2}{\includegraphics{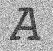}}\\[2mm]
      \end{minipage}}%
   \subfigure[]{ %
      \label{Fig:ocr:smooth}%
      \begin{minipage}[c]{0.3\linewidth}%
         \centering
         \scalebox{1.2}{\includegraphics{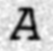}}\\[2mm]
      \end{minipage}}
   \subfigure[]{%
      \label{Fig:ocr:bin}%
      \begin{minipage}[c]{0.3\linewidth}%
      \centering
         \scalebox{1.2}{\includegraphics{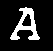}}\\[2mm]
      \end{minipage}}\\
   \subfigure[]{%
      \label{Fig:ocr:skl1}%
      \begin{minipage}[c]{0.4\linewidth}%
      \centering
         \scalebox{1.2}{\includegraphics{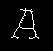}}\\[2mm]
      \end{minipage}}
   \subfigure[]{%
      \label{Fig:ocr:skl2}%
      \begin{minipage}[c]{0.4\linewidth}%
      \centering
         \scalebox{1.2}{\includegraphics{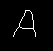}}\\[2mm]
      \end{minipage}}\\
\caption{%
Sample operations for shape recognition: (a) A noisy image; (b)
gaussian blur; and (c) segmentation by threshold. Shape width is
irrelevant for recognizing a letter, so we skeletonize the image (d),
and simplify it using a technique called ``multiscale pruning'' (e).
}\label{Fig:ocr}
\end{figure}

Let us show a sequence of operations to help the computer recognize
this letter. As the input, we have a noisy image, typically obtained
from cheap CCD's. We first Gaussian-blur~\cite{gonzalez} the image as to filter
out the low-scale noise and leave our letter `A'  in clear evidence:
\begin{verbatim}
 b = gsm2d(a,2);  // Gaussian smoothing with sigma = 2
 imshow(b,[]);    // the [] parameter rescales b to 0-1
\end{verbatim}

This is shown in figure~\ref{Fig:ocr:smooth}.  We may now perform thresholding with ease:
\begin{verbatim}
 c=im2bw(b,0.8); // threshold at the 80% level
 imshow(c,2);    // figure (c)
\end{verbatim}
as shown in figure~\ref{Fig:ocr:bin}. This a much simpler
representation for the letter A, with only two intensities. However,
we can make it even simpler. The width of the letter is of no
use to recognize it, so let us filter it out.  We do this using a
skeletonization algorithm, which finds a medial line of a given
shape. In SIP, there is a novel fast implementation of a useul skeletonization
algorithm~\cite{Estrozi:1999,Falcao:2002,Costa:2001} which is extremely easy to use:
\begin{verbatim}
 c = 1 - c;          // inverts the image
 t = skel(c);
\end{verbatim}

The  image `t' contains skeletons in every level of detail. We
select the level of detail for our application by thresholding
this variable:
\begin{verbatim}
 d = im2bw(t, 0.1);  // threshold at the 10% level
 imshow(d,2);        // figure (d)
\end{verbatim}

We now have a thin representation for `A' (figure~\ref{Fig:ocr:skl1}), but
there are unwanted spurs in the image. To filter them out, select a higher
threshold:
\begin{verbatim}
 d = im2bw(t,0.5);   // threshold at the 50% graylevel
 imshow(d,2);        // figure (e);
\end{verbatim}

This a clear-cut representation for the shape of the letter `A'
(figure~\ref{Fig:ocr:skl2}). It is now straightforward to make
some measures on this  shape in order to characterize it, a process
called ``feature extraction''~\cite{gonzalez}. The computer can
recognize the object as a letter A by doing statistics on these
measures or by applying techniques from Artificial Intelligence and Machine
Learning such as Neural Networks. Some measures useful to characterize the shape
is the number of holes, the width-to-height ratio, orientation of
the branch points, and more~\cite{Costa:2001}.

\section{Example: Segmentation of Blood Cells}~\label{Sec:blobs}
\begin{figure}[htb]
\centering
   \subfigure[]{ %
      \label{Fig:cells}%
      \begin{minipage}[c]{0.4\linewidth}%
         \centering
         \scalebox{0.6}{\includegraphics{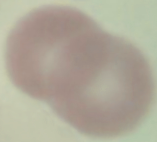}}\\[2mm]
      \end{minipage}}%
   \subfigure[]{ %
      \label{Fig:cells:bin}%
      \begin{minipage}[c]{0.4\linewidth}%
         \centering
         \scalebox{0.6}{\includegraphics{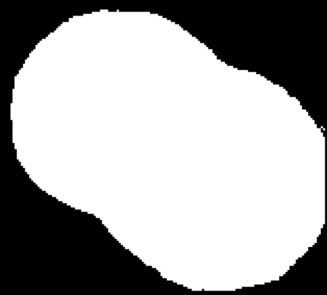}}\\[2mm]
      \end{minipage}}\\
   \subfigure[]{ %
      \label{Fig:cells:dt}%
      \begin{minipage}[c]{0.4\linewidth}%
      \centering
         \scalebox{0.33}{\includegraphics{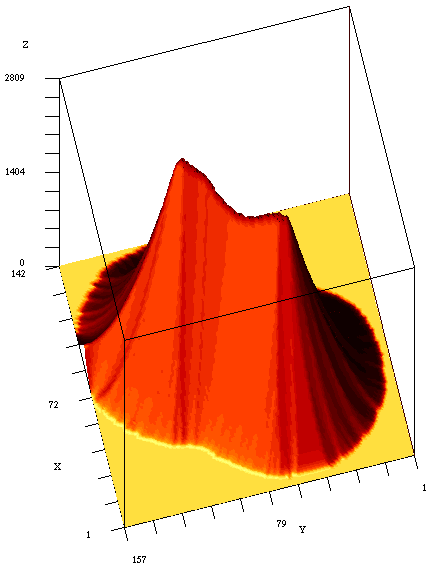}}\\[2mm]
      \end{minipage}}%
   \subfigure[]{%
      \label{Fig:cells:separated}%
      \begin{minipage}[c]{0.4\linewidth}%
      \centering
         \scalebox{0.6}{\includegraphics{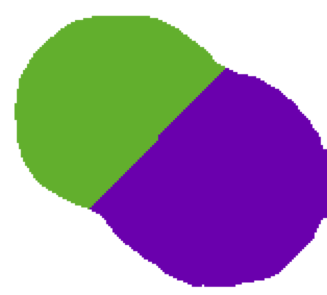}}\\[2mm]
      \end{minipage}}
\caption{%
Separating overlapping blood cells, shown in (a).  Steps: (b)
initial segmentation; (c) euclidean distance transform; and (d)
correctly segmented image.
}\label{Fig:blood}
\end{figure}

Another task frequently found when solving practical problems is
the automatic separation of overlapping objects. Suppose we have
an image of many round or oval objects, such the a microscope
image of blood cells. If we want to count the number of cells,
there are problems with cases such as in Figure~\ref{Fig:blood},
with two overlying blood cells. They must count not as one, but two.

The input image is downloadable at:
\begin{center}
  \texttt{http://siptoolbox.sourceforge.net/tutorial/images/cells.png}
\end{center}

Start by reading the image and threshold it:
\begin{verbatim}
 a = gray_imread('cells.png');  // reads in grayscale
 imshow(a);
 b = im2bw(a, 0.9);
 imshow(b,2);
\end{verbatim}

Figure~\ref{Fig:cells:bin} shows the result. We now calculate the
distance transform of the inverse image:
\begin{verbatim}
 b = 1 - b;     // inverts image
 d = bwdist(b); // distance transform (euclidean distances)
 d = sqrt(d);   // square root of the values
 d = normal(d); // rescales to 0-1 range
\end{verbatim}

To understand the distance transform, we use a command
from the ENRICO toolbox to make a nice 3D plot:
\begin{verbatim}
 set(gcf(), 'auto_clear','off'); //disables automatic clear
 [nrows, ncols] = size(d);
 figure(2); // use another window
 setcmap();        // chose the 'Hot' colormap
 set(gcf(),'pixmap',1);  // double-buffering mode (faster)
 shadesurf2(1:nrows, 1:ncols, d *200); // comp. the surface
 xset('wshow');  // shows the surface
\end{verbatim}

You may click on the button '3D Rot.'\ to rotate the surface.
The above commands render the distance transform as a surface
(figure~\ref{Fig:cells:dt}).
The higher the surface, the greater the distance of the corresponding
pixel to the image background. Note that the peaks of the distance 
transform are in the middle of each blob. The idea is to run Watershed 
segmentation~\cite{Soille:1991} using those peaks as
markers. For this, complement the distance transform so that the peaks become minima:
\begin{verbatim}
d = 1-d;
\end{verbatim}

Apply a slight median filter~\cite{gonzalez} to eliminate spurious minima:
\begin{verbatim}
dm = mogrify(d,['-median','2']);
\end{verbatim}

Force pixels that don't belong to the objects to be at zero distance:
\begin{verbatim}
dm = dm .* b;
\end{verbatim}

And finally perform watershed segmentation:
\begin{verbatim}
w = watershed(dm);
clf;
imshow(w, rand(10,3));  // use 10 colors at most
\end{verbatim}

Variable 'w' is an image with a unique number for each watershed
region.  The imshow with a random colormap displays each region
with a unique arbitrary color. Note how the regions were correctly
separated by watershed, except for the hardest cases.  We can use
this procedure for automatic counting objects in an image with
many cells, and it's robust to overlaps.  This is specially
useful to deal with bigger images at a large scale.

If the image had many cells, by the same procedure it would be
straightforward to count the regions:
\begin{verbatim}
 n = maxi(w) - 1;   // subtract 1 for background
\end{verbatim}

\section{Conclusions}
Scilab and SIP constitute a powerful and easy-to-use environment for
rapid development of image processing solutions. The expressiveness
of the Scilab interpreted language allows image manipulation
in a natural way. Together with the SIP toolbox, we showed how
simple it is to carry out complex operations such as the FFT,
watershed segmentation, and multiscale skeletonization, in this environment.

As authors of the SIP Toolbox, we attest that the free
software~\footnote{free as in ``free speech''; also called 'libre'.}
model -- an Open Source model~\cite{Raymond:2001} which is guaranteed to
always remain free -- has been essential to its development
and success.  First of all, the wide visibility of any free software
easily sustains its development after once it is born. The feedback
provided by direct contact with users and many co-developers also
copes for the stability and usability of the toolbox.  Furthermore,
all the third-party packages used by SIP are Open Source: external
libraries such as ImageMagick, Animal and OpenCV, as well as Scilab
itself. This provides an excellent development environment, where
any bugs in the third-party components may be reported directly
to their authors, or even fixed by ourselves without bureaucracy.
In addition, the community benefits a lot.  Many scientists have
replaced expensive proprietary software even big companies can't afford. They
also benefit from the sourcecode of many algorithms, by studying, evaluating and
comparing them.

There is a clear demand in the Computer Vision community for
software packages like SIP.  Currently there is no other free
software that provides easy scripting for image analysis together
with the advanced techniques available in SIP. The number of supported
file formats supersedes many proprietary software, such as Matlab.
As a consequence, many people in universities and some industries have been
using SIP, specially researchers and Image Processing students. The
toolbox has reportedly been used in countries such as Germany, USA,
Brazil, Greece, and France.


\end{document}